\documentclass[11pt]{article}
\usepackage{moriond,epsfig}

\def\be{\begin{equation}}
\def\ee{\end{equation}}
\def\bea{\begin{eqnarray}}
\def\eea{\end{eqnarray}}

\begin{document}
\vspace*{0cm}
\title{LATEST RESULTS FROM THE MiniBooNE EXPERIMENT AND UPDATED $\nu_{\mu} \rightarrow \nu_e$ SENSITIVITY}

\author{ J. Monroe for the MiniBooNE collaboration}

\address{Columbia University Department of Physics, \\ 
538 W. 120th St., New York, NY, 10027, USA}

\maketitle\abstracts{
Neutrino oscillations have been established in solar and atmospheric neutrinos, but a third signal from the LSND experiment is incompatible with three Standard Model neutrinos.  The MiniBooNE experiment can confirm or refute the LSND oscillation signal with $1 \times 10^{21}$ protons on target.  Preliminary results on $\nu_{\mu}$ analyses and an updated $\nu_{\mu} \rightarrow \nu_e$ sensitivity are presented.}

\section{Introduction}
Neutrino masses have been conclusively established via positive neutrino oscillation measurements of solar and atmospheric neutrinos \cite{solar}$^,$ \cite{atmospheric}.  A third and unconfirmed positive result comes from the LSND experiment, a short baseline accelerator $\overline{\nu}_{\mu} \rightarrow \overline{\nu}_e$ oscillation search \cite{lsndfinal}.  The MiniBooNE experiment at Fermi National Accelerator Laboratory will confirm or refute the LSND result with higher statistics and different sources of systematic error.  

The best fit values for LSND, atmospheric, and solar oscillations are $\Delta m^2_{LSND} \sim 0.3 - 3 \ eV^2$, $\Delta m^2_{atmospheric} \sim 2.5 \times 10^{-3} \ eV^2$, and $\Delta m^2_{solar} \sim 7 \times 10^{-5} \ eV^2$.  These neutrino oscillation parameters cannot be accommodated within the Standard Model framework of three weakly interacting neutrinos since three neutrinos allow only two independent values of $\Delta m^2$.  If MiniBooNE confirms the oscillation interpretation of the LSND result, new physics will be required to accommodate the solar, atmospheric, and LSND signals.  A number of theoretical models have been proposed to explain the experimental results \cite{plausibletheories}.   

\begin{figure}
\mbox
{
\begin{minipage}{0.5\textwidth}
\psfig{figure=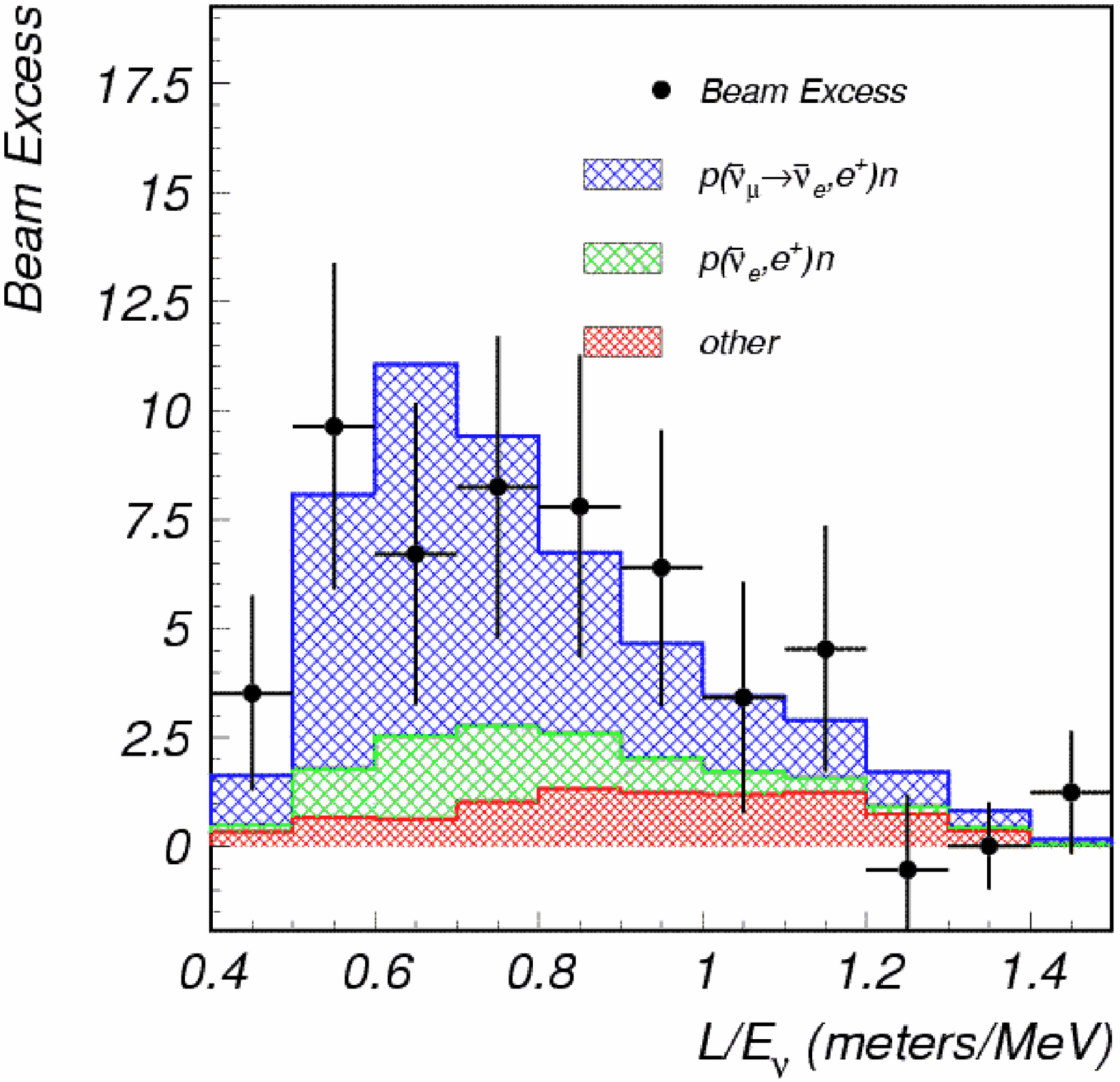,height=2.5in}
\caption{LSND data excess vs. (L/E) ($meters/MeV$), with backgrounds and best-fit oscillation hypothesis.\label{fig:LSNDresult1}}
\end{minipage}
\hspace{0.2cm}
\begin{minipage}{0.5\textwidth}
\psfig{figure=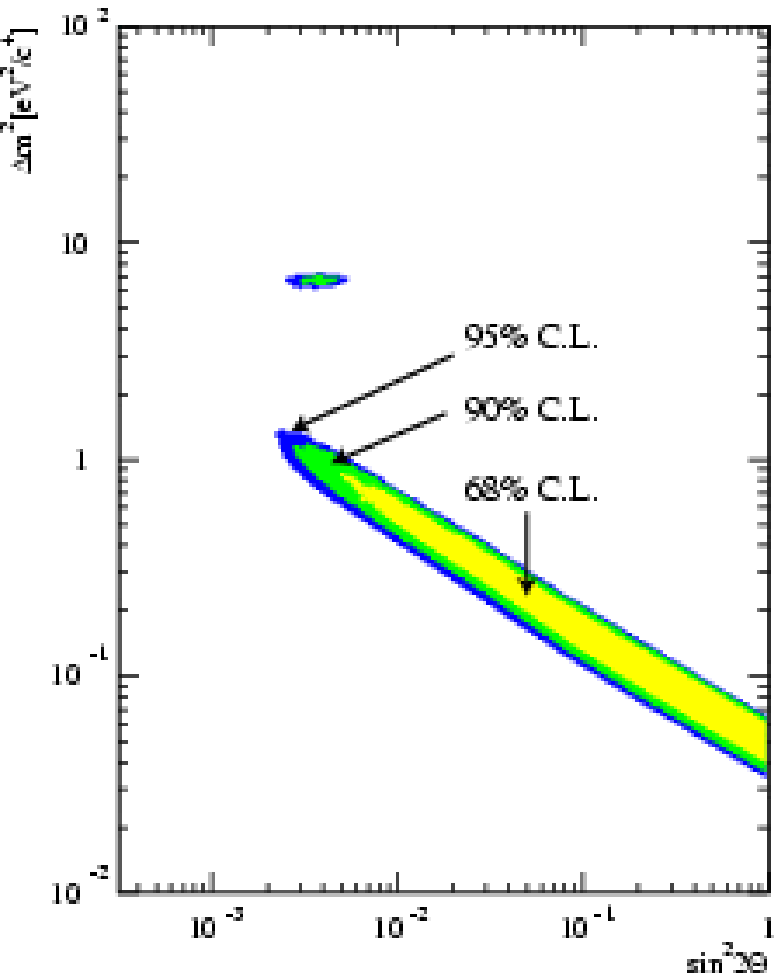,height=2.5in}
\caption{$\Delta m^2$ vs. $\sin^2 2\theta$ for joint analysis of KARMEN2 and LSND data.\label{fig:LSNDresult2}}
\end{minipage}
}
\end{figure}

The LSND experiment, which ran from 1992 - 1997 at Los Alamos National Laboratory, observed an excess of 87.9 $\pm$ 22.4 $\pm$ 6.0 $\overline{\nu}_e$ events in the almost pure $\overline{\nu}_{\mu}$ beam from $\pi$ decay at rest.  The statistical significance of the excess is 3.8$\sigma$.  The mean beam energy was $\sim$30 $MeV$ and the distance from the neutrino source to the detector was $\sim$30 $m$.  The beam-off background subtracted $\overline{\nu}_e$ excess with the best-fit oscillation hypothesis is shown as a function of $L/E$ in figure \ref{fig:LSNDresult1}.  The significance of the oscillation hypothesis is slightly lower, 3.3$\sigma$, due to the inclusion of additional systematic errors on the determination of the beam neutrino flux.  The oscillation probability measured for $\overline{\nu}_{\mu}$ from $\pi$ decay at rest was 0.264 $\pm$ 0.067 $\pm$ 0.045 \%.  The KARMEN2 experiment had a similar $L/E$ to LSND, and observed no oscillation signal \cite{karmenfinal}.  A joint analysis of the LSND and KARMEN2 data was performed by collaborators from both experiments, and the resulting allowed regions are shown in figure \ref{fig:LSNDresult2} \cite{joint}.
\section{MiniBooNE Experiment Overview}
The MiniBooNE experiment is designed to confirm or refute definitively the LSND oscillation signal.  The MiniBooNE detector is located 541 $m$ from a neutrino source on the Booster neutrino beam line at Fermi National Accelerator Laboratory.  The mean neutrino energy is $\sim 700$ $MeV$; therefore the $L/E$ is similar to that of LSND, which gives MiniBooNE sensitivity to the same range of $\Delta m^2$ values.
\subsection{Neutrino Beam}
The MiniBooNE neutrino beam is produced from 8.89 $GeV/c$ protons incident on a thick beryllium target located inside a magnetic focusing horn.  Typical operating conditions are $4 \times 10^{12}$ protons per pulse, at 3 - 4 $Hz$, with a beam uptime of $\sim 88\%$.  The beam spill duration is 1.6 $\mu s$.  The focusing horn increases the neutrino flux at the detector by a factor of $\sim$5, and can operate in both positive and negative polarities for $\nu$ and $\overline{\nu}$ running; currently positive sign mesons are selected.  Mesons produced in the target decay in a 50 $m$ long decay pipe.  The neutrino beam is primarily composed of $\nu_{\mu}$ from $\pi^+$ decay, with a 0.5\% contamination from $\nu_e$.  The processes that contribute to the intrinsic $\nu_e$ in the beam are $\mu^+ \rightarrow e^+ \nu_e \overline{\nu}_{\mu}$, $K^+ \rightarrow \pi^0 e^+ \nu_e$, and $K^0_L \rightarrow \pi^{\pm} e^{\mp} \nu_e$.  The neutrino flux is shown in figure \ref{fig:mboone} (left).
\subsection{Detector \label{DETsection}}
The MiniBooNE detector is a 6.1 $m$ radius sphere filled with mineral oil ($CH_2$).  There are 1280 inward-facing ``tank'' PMTs, and  240 outward-facing ``veto'' PMTs.  Particle identification depends upon both prompt Cerenkov and time-delayed scintillation light.  Neutrino induced events are identified by requiring that the event occur within the beam spill, have fewer than 6 veto PMT hits, and have greater than 200 tank PMT hits.  With these simple cuts, the cosmic ray background rejection is $>$999:1000.  A fiducial volume cut at $R < 5 \ m$ is also typically required to ensure good energy reconstruction.  

\begin{figure}[ht]
\hfill
\begin{minipage}{2.0in}
\epsfxsize=2.0in\epsfbox{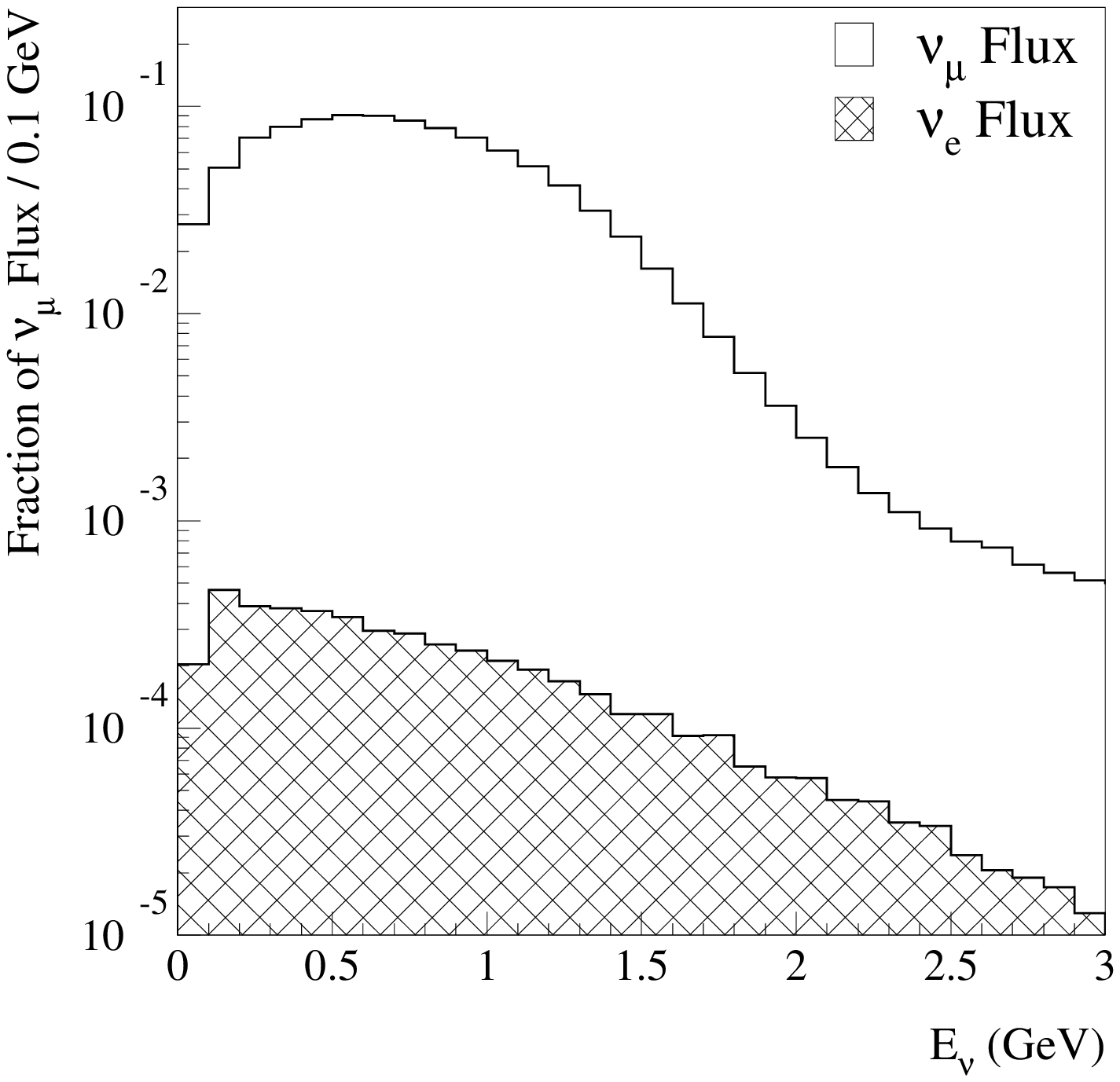}
\end{minipage}
\hfill
\begin{minipage}{2.0in}
\epsfxsize=2.0in\epsfbox{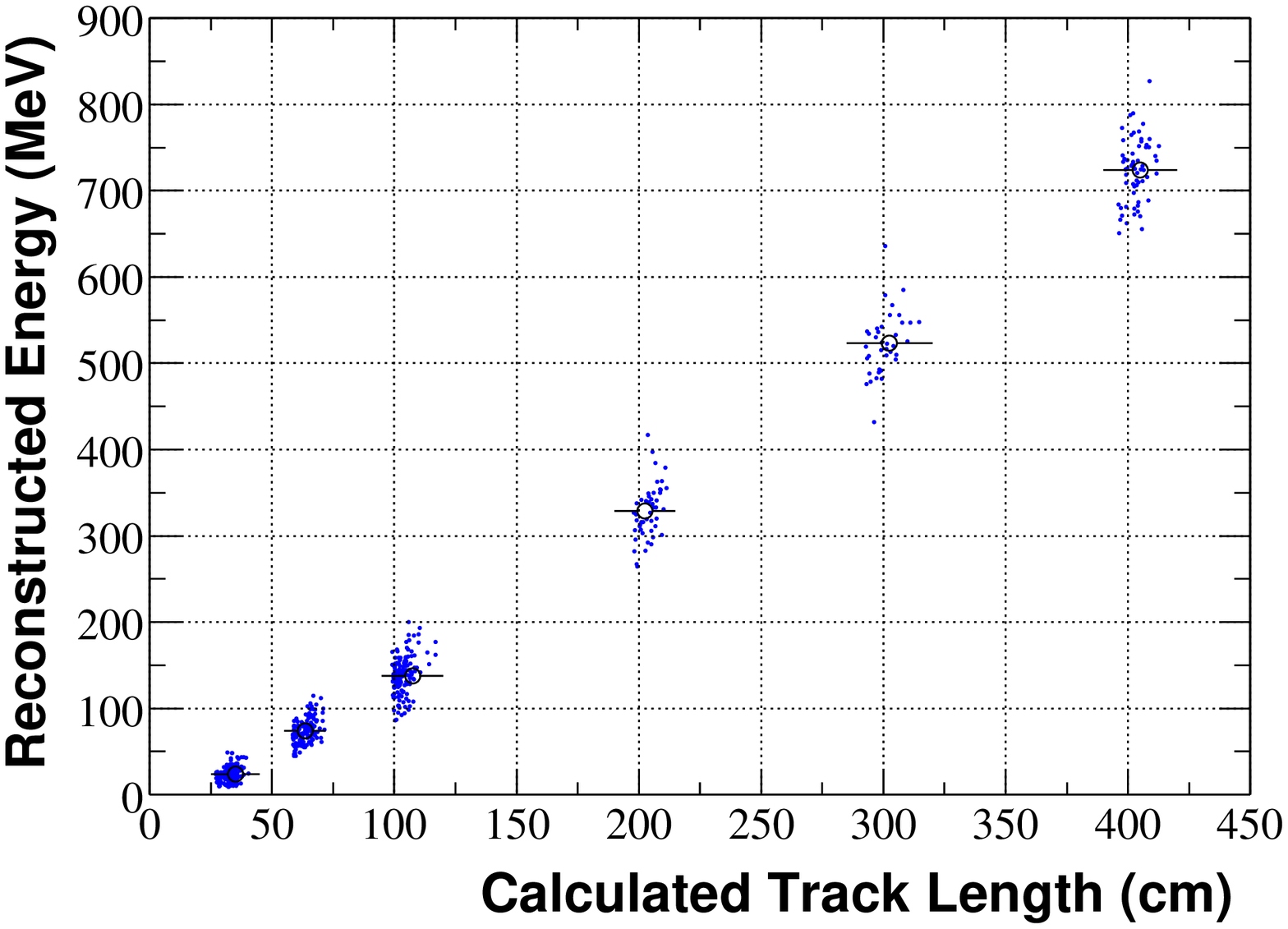}
\end{minipage}
\hfill
\begin{minipage}{2.0in}
\epsfxsize=2.0in\epsfbox{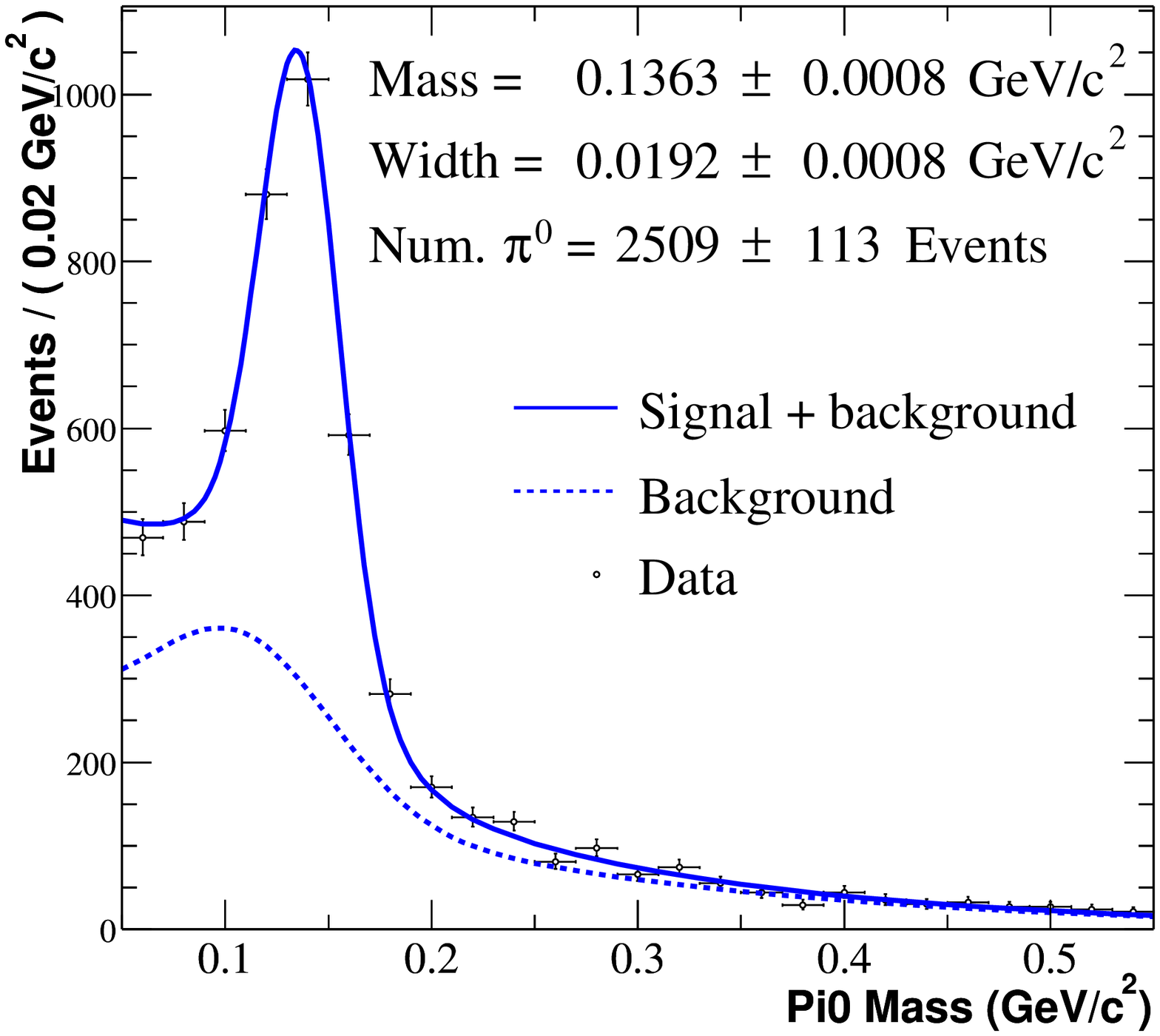}
\end{minipage}
\caption{Left: Preliminary MiniBooNE neutrino flux vs. $E_{\nu}$ ($GeV$).  Middle: Preliminary stopping muon data reconstructed energy vs. calculated track length.  Right: Preliminary $\pi^0$ data reconstructed $\pi^0$ mass peak.\label{fig:mboone}}
\end{figure}

The calibration relies on four primary components: a laser system, stopping cosmic ray muons, Michel electrons from muon decay, and photons from $\pi^0$ decays.  The laser system is used to calibrate continuously the charge and time response of the tank PMTs.  Stopping cosmic ray muons are used to calibrate the energy scale for muon-type events, and measure the position and angle reconstruction resolution.  The entering position and angle are tagged with a hodoscope situated above the detector, and the stopping position can be measured if the muon decays in one of six small scintillator cubes deployed within the detector volume.  The calculated track length from the muon tracker is shown vs. reconstructed energy in figure \ref{fig:mboone} (middle).  Michel electrons from muon decay are used to calibrate the energy scale and measure the energy reconstruction resolution of electron-type events at the endpoint of the Michel decay spectrum, currently $\sim$12\% at 52.8 $MeV$.  Photons from $\pi^0$ decays are also used to calibrate the energy scale of electron-type events, over a larger energy range (0.04 - $\sim$1 $GeV$).  The reconstructed $\pi^0$ mass is shown in figure \ref{fig:mboone} (right).
\section{$\nu_{\mu}$ Events}
The event rate prediction is based on the product of neutrino flux and cross section.  The neutrino flux is modeled with a GEANT4 based Monte Carlo \cite{G4} and an external parameterization of the $p \ Be \ \rightarrow \ \pi^+ \ X$ cross section.  The parametrization comes from a global fit, assuming the Sanford-Wang model \cite{SW}, to the world's data for this reaction in the range $6 \ GeV/c \ < \ p \ < \ 17 \ GeV/c$.  The NUANCE Monte Carlo \cite{nuance} is used to predict the neutrino interaction cross section.  NUANCE uses standard models to simulate the various $\nu$ interaction processes on CH2, which include the Llewellyn-Smith free nucleon quasi-elastic cross section with $m_A \ = \ 1.03 \ GeV/c^2$, the Rein-Sehgal resonance and coherent pion production cross section, a Smith-Moniz Fermi gas model, and final state interactions based on $\pi$ - Carbon scattering data.  

At MiniBooNE neutrino energies, the cross section has contributions from several processes including charged current quasi-elastic scattering (39\% of the total event rate), charged current resonance production (25\%), neutral current elastic scattering (7\%), and neutral current $\pi^0$ production (7\%).  For the $\nu_{\mu} \rightarrow \nu_e$ oscillation analysis, the most important processes are CCQE scattering which enables a precise measurement of the neutrino energy, NC $\pi^0$ production which is a large background to a $\nu_e$ signal, and NC elastic scattering which can be used to study the optical properties of the detector.  The MiniBooNE detector Monte Carlo and reconstruction are tuned on calibration data only; so to verify the Monte Carlo predictions, these three classes of $\nu_{\mu}$ events are currently under investigation.
\subsection{CCQE Events \label{CCQEsection}}
\begin{figure}[ht]
\hfill
\begin{minipage}{2.0in}
\epsfxsize=2.0in\epsfbox{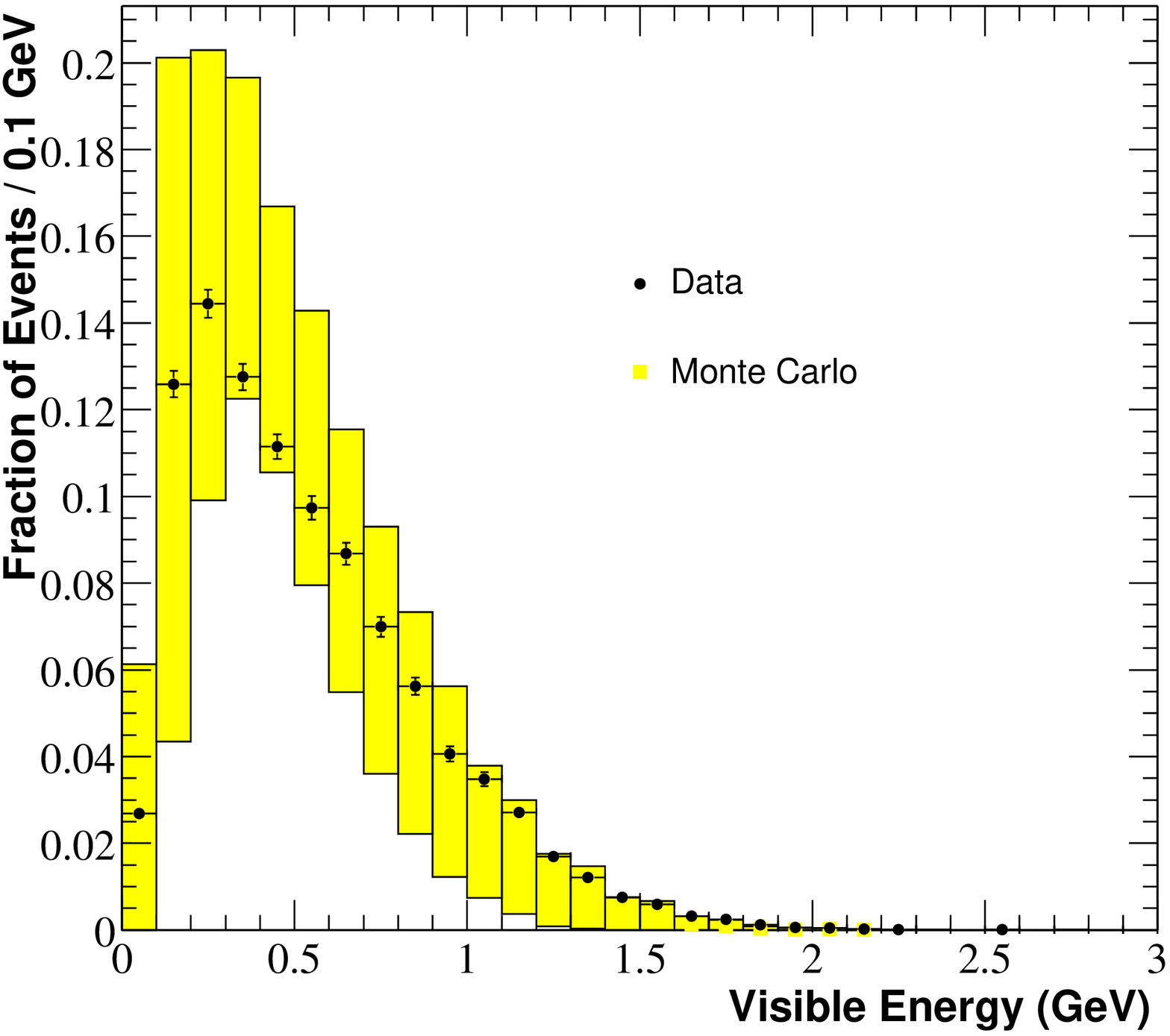}
\end{minipage}
\hfill
\begin{minipage}{2.0in}
\epsfxsize=2.0in\epsfbox{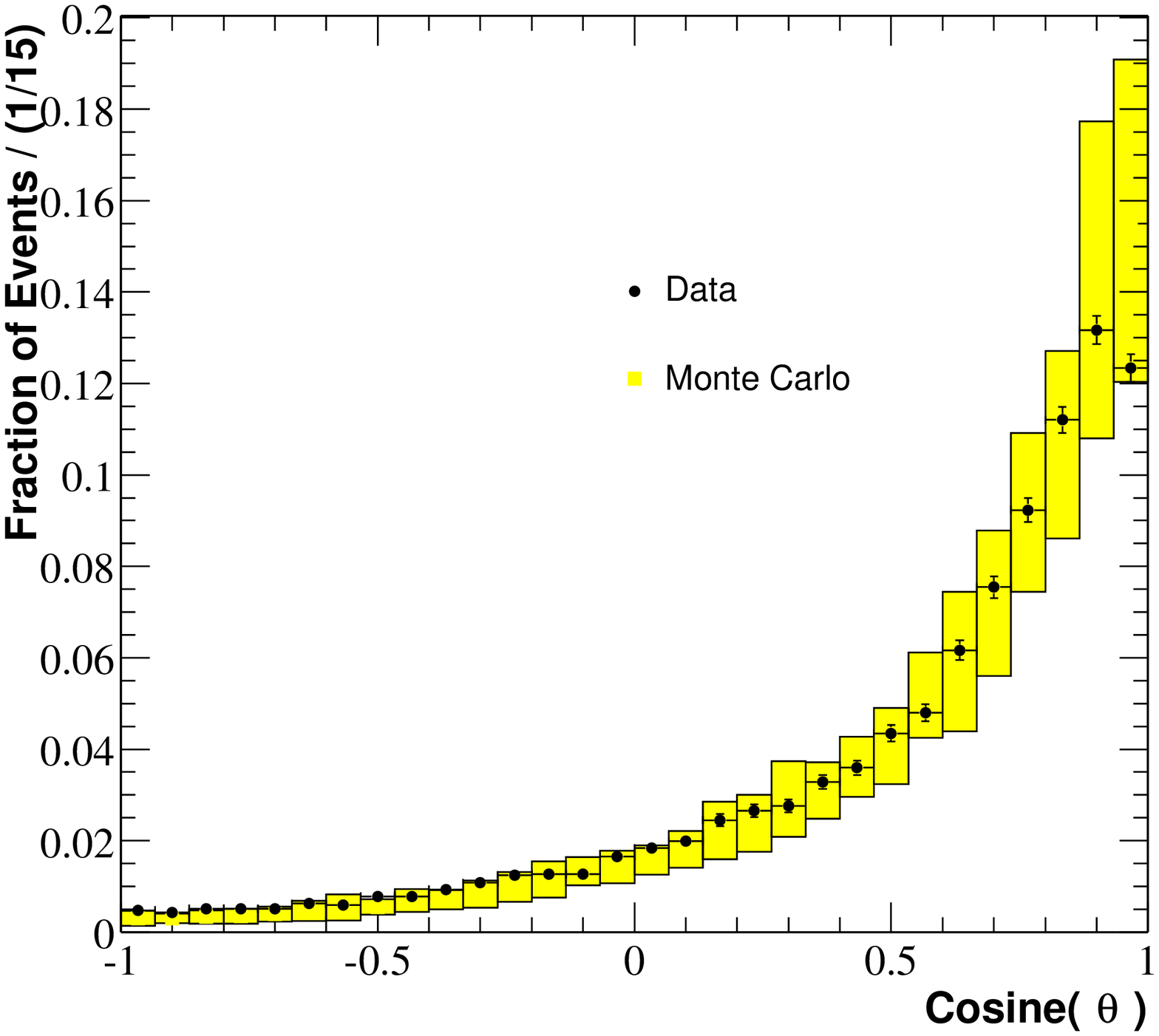}
\end{minipage}
\hfill
\begin{minipage}{2.0in}
\epsfxsize=2.0in\epsfbox{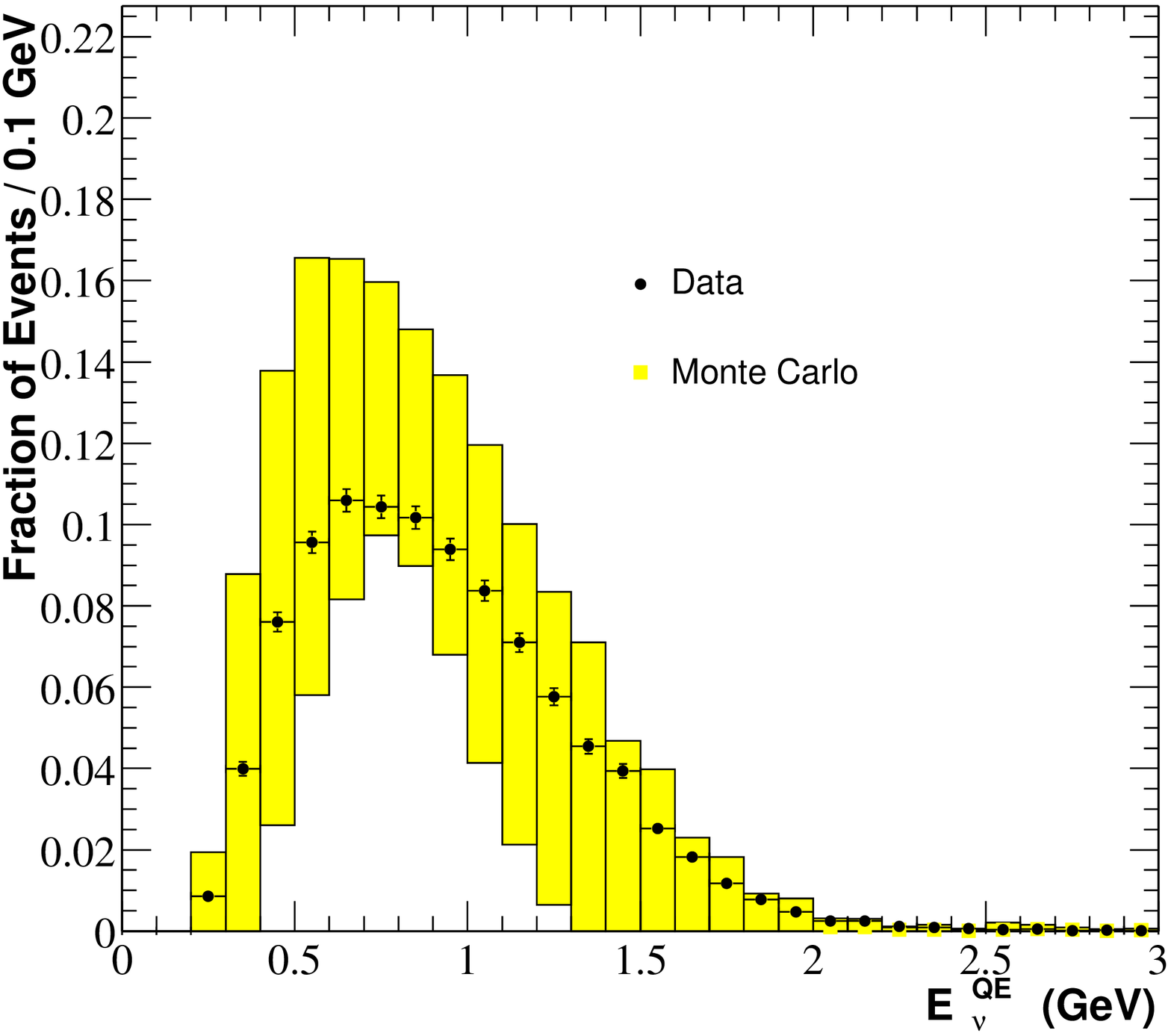}
\end{minipage}
\caption{Preliminary data and Monte Carlo reconstructed visible energy of $\nu_{\mu}$ CCQE events (left), Preliminary reconstructed $\cos(\theta_{beam})$ (middle), and Preliminary reconstructed $E_{\nu}^{QE}$ (right).  All distributions are normalized to unit area, error bars on data are statistical only, and Monte Carlo error bars include flux, cross section, and oil optical model variations.\label{fig:CCQE}}
\end{figure}

Quasi-elastic kinematics enable a precise determination of the neutrino energy in $\nu_{\mu} (\nu_e) \ n \rightarrow \ \mu^- (e^-) \ p$ interactions.  Neglecting corrections for the motion of the target nucleon, the neutrino energy can be calculated with only the measured energy and angle of the final state muon, which are shown with the Monte Carlo expectation in figure \ref{fig:CCQE} (left and middle).  The reconstructed neutrino energy is shown in figure \ref{fig:CCQE} (right).  The current neutrino energy resolution is $\sim$15 - 20\% and is expected to be $\sim$10\% at $E_{\nu}$ = 1 $GeV$ as background rejection improves.

The event selection requires that the event pass the cosmic background and fiducial volume cuts described in section \ref{DETsection}, and that the event topology be consistent with one muon-induced Cerenkov ring.  The efficiency of the event selection is 30\%, and results in an 88\% pure $\nu_{\mu}$ CCQE data set. The background is almost entirely due to charged current single pion production events, $\nu_{\mu} \ p \ \rightarrow \ \mu^- \ p \ \pi^+$ where the $\pi^+$ is produced below its Cerenkov threshold.  The current $\nu_{\mu}$ CCQE data set comprises $\sim$60,000 events, and will also be used to search for $\nu_{\mu} \rightarrow \nu_{sterile}$ oscillations. 
\subsection{Neutral Current $\pi^{0}$ Events}
Understanding the rate and kinematics of neutral current $\pi^{0}$ production is critical to the $\nu_{e}$ oscillation analysis since these events represent a large background.  The event selection requires that the events pass the cosmic rejection and fiducial volume cuts described in section \ref{DETsection}, and that each $\gamma$ from $\pi^0$ decay have $>$ 40 $MeV$ to ensure good reconstruction.  Signal events are extracted on a statistical basis using fitted curves from Monte Carlo-based parameterizations, where the signal contribution arises from NC resonant and coherent single $\pi^{0}$ events, and the background contribution comes from all other events (which includes single and multi-$\pi^0$ final states), shown in figure \ref{fig:mboone} (right).  The current data sample has $2425\pm107$ signal-like NC $\pi^{0}$ events after signal extraction.

\begin{figure}[ht]
\hfill
\begin{minipage}{2.0in}
\epsfxsize=2.0in\epsfbox{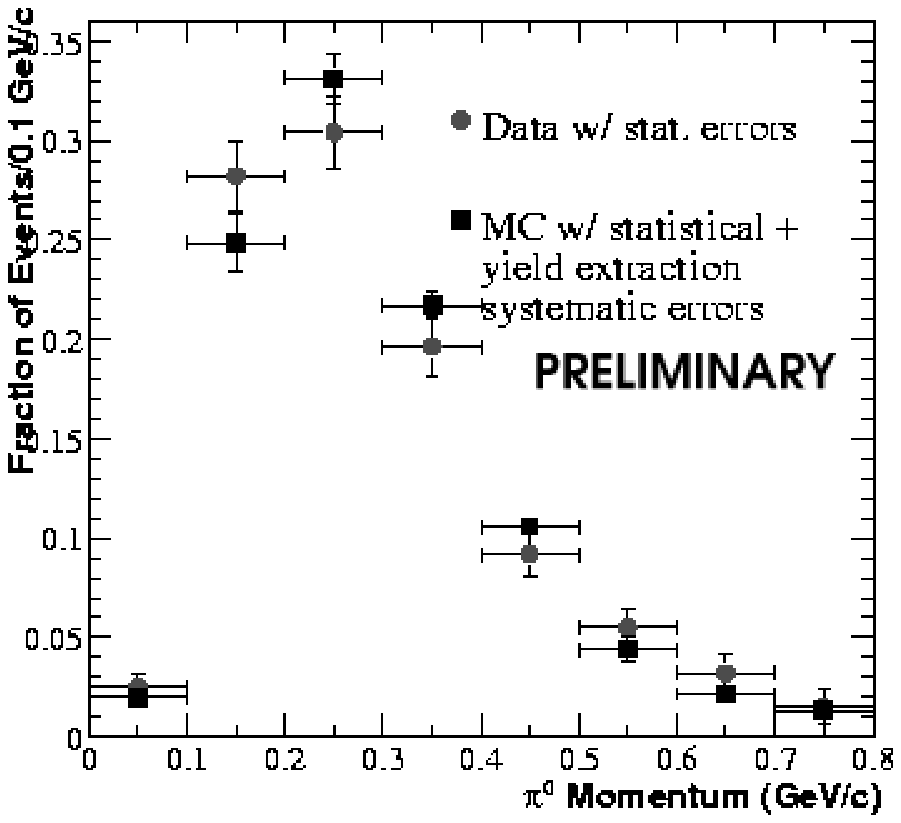}
\end{minipage}
\hfill
\begin{minipage}{2.0in}
\epsfxsize=2.0in\epsfbox{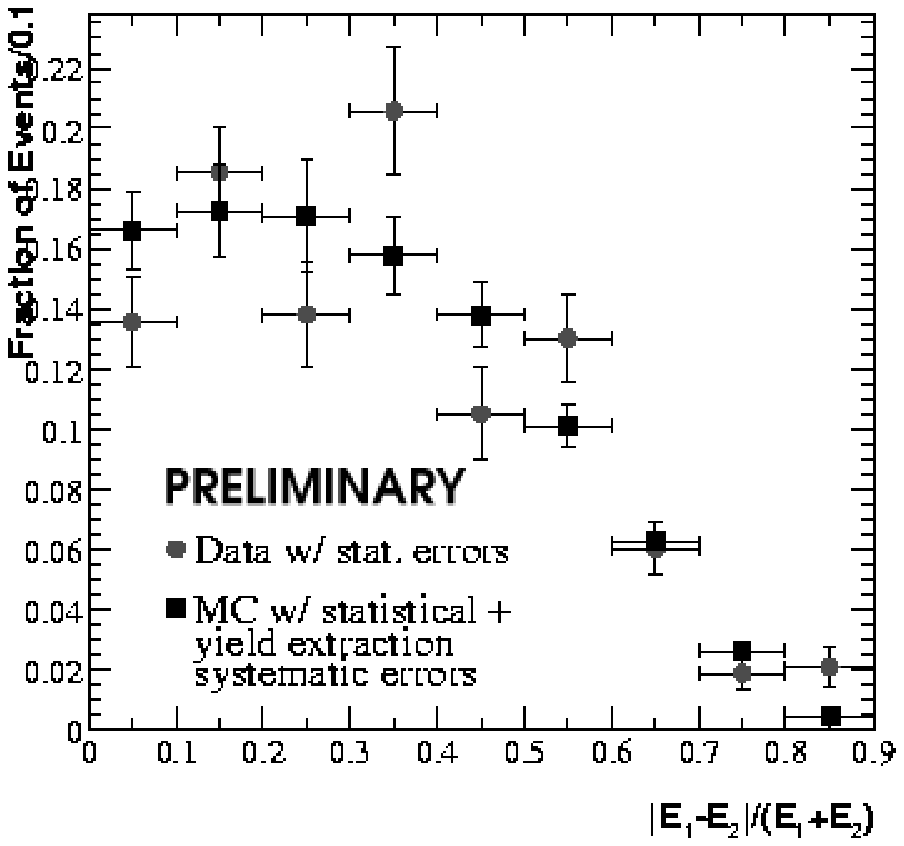}
\end{minipage}
\hfill
\begin{minipage}{2.0in}
\epsfxsize=2.0in\epsfbox{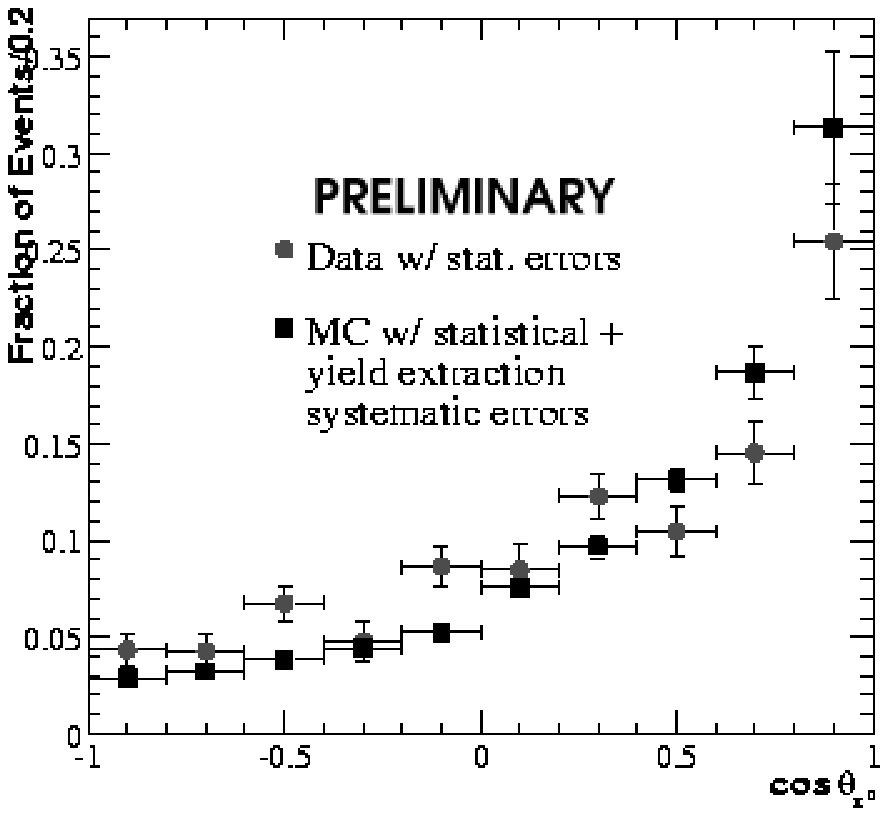}
\end{minipage}
\caption{Data and Monte Carlo extracted $\pi^{0}$ yields in $p_{\pi^{0}}$ (left), energy asymmetry (middle), and $\cos~\theta_{\pi^{0}}$ (right).  All distributions are normalized to unit area, error bars on data are statistical only, and Monte Carlo error bars are estimated systematic errors from the yield extraction procedure, added in quadrature with statistical errors.\label{fig:NCpi0}}
\end{figure}

The number of NC $\pi^{0}$'s is also extracted \footnote{The signal fraction in each bin is extracted via a fit to the invariant mass plot for events in that bin.} in bins of $p_{\pi^{0}}$, $\cos(\theta_{\pi^{0}})$, and $\frac{|E_{1}-E_{2}|}{E_{1}+E_{2}}$.  The distributions are compared with the Monte Carlo expectation in figure \ref{fig:NCpi0}.  The energy asymmetry of the $\gamma$'s from $\pi^{0}$ decay is important since highly asymmetric decays mimic the $\nu_e$ signal \footnote{The energy asymmetry of the $\gamma$'s from $\pi^{0}$ decay falls off just above 0.6 due to the minimum energy requirement for each $\gamma$.}. The $\cos(\theta_{\pi^{0}})$ distribution is sensitive to the production mechanism and will be used to determine the contribution of coherent production.  This result will have interesting implications for the Super-K analysis which uses NC $\pi^0$ events to differentiate $\nu_{\tau}$ from $\nu_{sterile}$ events.  
\subsection{Neutral Current Elastic Scattering Events}
Neutral current elastic interactions produce low energy nucleons, and a relatively high amount of scintillation light in the detector since the final state has only nuclear fragments.  These events provide an essential cross-check of the oil optical model, and test the quality of the reconstruction for low energy events.

\begin{figure}[ht]
\centerline{\epsfxsize=6.5in\epsfbox{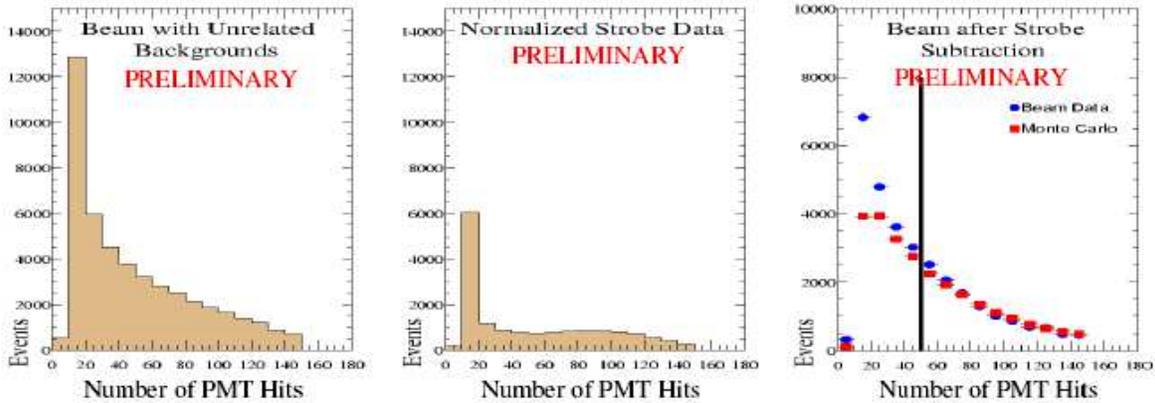}}
\caption{\label{fig:NCE_answer} NC elastic beam-on data tank hits (left), beam-off data tank hits (middle), Monte Carlo and data with strobe subtraction, relatively normalized.  50 tank hits $\simeq$ 150 $MeV$ proton kinetic energy.}
\end{figure}

The primary background to NC elastic events is cosmic muons, and the subsequent Michel electrons.  A beam-off background subtraction \footnote{Random strobe triggers which contain the beam-unrelated backgrounds from cosmics and environmental radioactivity are subtracted from beam-on data.} is used to eliminate these events on a statistical basis.  The event selection further requires that the event pass the cosmic ray background rejection cuts described in section \ref{DETsection} and that it have $<$150 tank PMT hits.  Figure \ref{fig:NCE_answer} shows the subtracted distribution of tank hits for events that pass the NC elastic selection cuts.  The efficiency and purity are 60\% and 81\% respectively.  With the current data set, there are $\sim$25,000 events.
\section{Updated Oscillation Sensitivities}
\begin{figure}
\mbox
{
\begin{minipage}{0.5\textwidth}
\hspace{0.5cm}
\psfig{figure=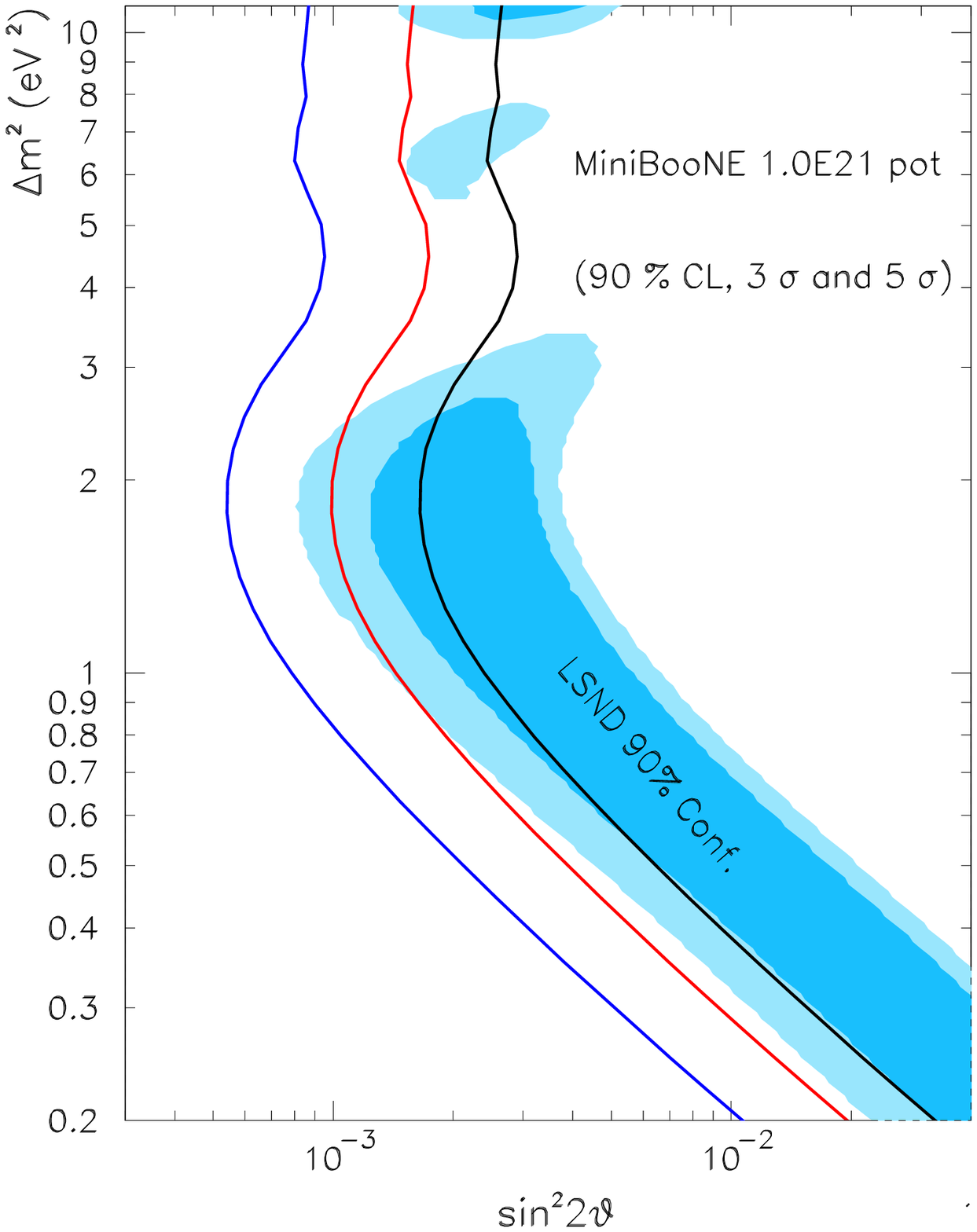,height=3.0in}
\caption{MiniBooNE $\nu_{\mu} \rightarrow \nu_e$ sensitivity at 90\% C.L., 3$\sigma$, and 5$\sigma$ coverage of the LSND allowed region.\label{fig:MBsens1}}
\end{minipage}
\hspace{0.2cm}
\begin{minipage}{0.5\textwidth}
\hspace{0.5cm}
\psfig{figure=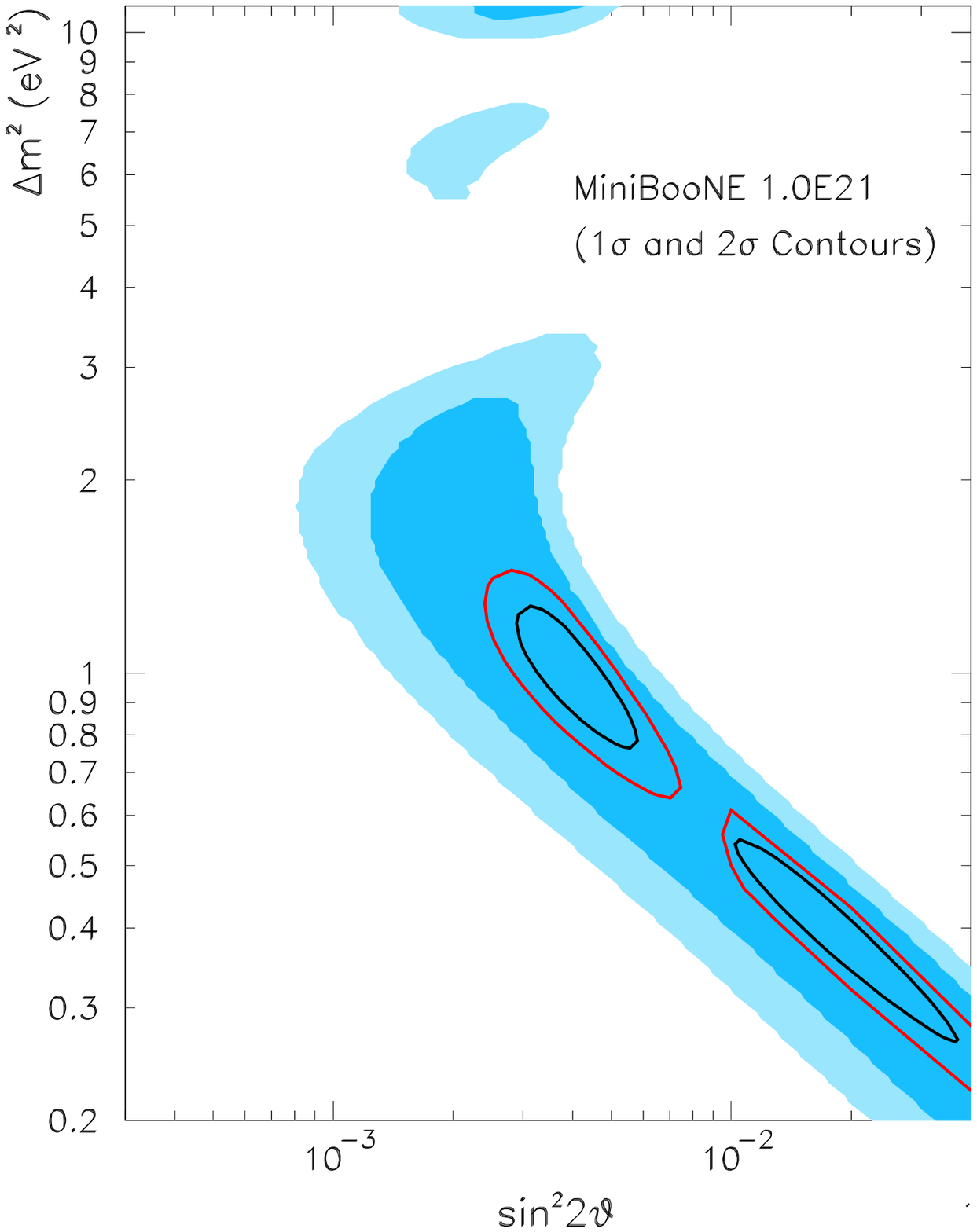,height=3.0in}
\caption{MiniBooNE $\nu_{\mu} \rightarrow \nu_e$ parameter measurement capability at 1$\sigma$ and 2$\sigma$ for high and low $\Delta m^2$.\label{fig:MBsens2}}
\end{minipage}
}
\end{figure}

The oscillation sensitivity has been updated with Monte Carlo tuned on calibration data, current reconstruction, particle ID, event selection, and event rate predictions taken from the data normalization to protons on target.  In Monte Carlo, the number of signal $\nu_e$ CCQE events is $\sim$300, assuming the average LSND oscillation probability, with $\sim$350 background events from mis-identified $\nu_{\mu}$-induced $\mu^+$ and $\pi^0$, and $\sim$350 background events from $\nu_e$ in the beam from $\mu^+$, $K^{+}$, and $K^0_L$ decays.  The systematic errors on the backgrounds are assumed to be 5 - 10\%.  The $\nu_{\mu} \rightarrow \nu_e$ sensitivity for $1 \times 10^{21}$ protons on target is shown in figure \ref{fig:MBsens1}, and the oscillation parameter measurement capability in the event of a positive signal at high or low $\Delta m^2$ is shown in figure \ref{fig:MBsens2}.   
\section{Conclusions}
The MiniBooNE experiment can conclusively confirm or refute the LSND oscillation signal with $1 \times 10^{21}$ protons on target.  To date, $\sim$20\% of the expected protons have been collected.  Analyses of $\nu_{\mu}$ charged current quasi-elastic, neutral current $\pi^0$, and neutral current elastic data sets are well underway, and the $\nu_{\mu} \rightarrow \nu_e$ analysis is on track for results in 2005.

\section*{Acknowledgments}
MiniBooNE gratefully acknowledges the support it receives from the Department of Energy and from the National Science Foundation.  The presenter of this paper was supported by NSF grant PHY-98-13383.

\end{document}